# Biaxial nematics with $C_{2h}$ symmetry composed of calamitic particles. A molecular field theory

Rauzah Hashim,[a] Geoffrey R. Luckhurst[b*] and Hock-Seng Nguan(阮福成)[a,b]

[a]*Department of Chemistry, University of Malaya, 50603 Kuala Lumpur, Malaysia.*

[b] *School of Chemistry, University of Southampton, Highfield, Southampton, SO17 1BJ, United Kingdom.*

PACS number(s): 61.30. Gd,64.70.M-, 61.30.Cz

A molecular field theory of biaxial nematics formed by molecules with $C_{2h}$ point group symmetry has been developed by Luckhurst *et al*. and a Monte Carlo computer simulation study of this model has been performed by Hashim *et al.*. In these studies the truncated model pair potential was only applied to molecules whose long axes are taken to be along their $C_2$ rotation axes. The present study extends this work by assuming that the molecular long axis is now perpendicular to the $C_2$ axis, resulting in there being two possible choices of minor axes. It considers the phases formed by both cases. The molecular field theory for these models is formulated and reported here. The theoretical treatment of the present cases gives rise to a new set of order parameters. So as to simplify the pseudo-potentials only the dominant second rank order parameters are considered and evaluated to give the phase behaviour of these truncated models. The predicted phase behaviour is compared with the results from the molecular field study of the previous model potential.

## I. INTRODUCTION

The first molecular field approach to nematic liquid crystals was given by Grandjean [1] in 1917, however this work went largely unnoticed [2]. Many years later, in 1958, Maier and Saupe [3-5] developed a more detailed and complete molecular field theory. Since then the Maier-Saupe molecular field approach for the uniaxial nematic phase remains the most valuable and widely used. Their theory started from a simple pair-interaction for calamitic molecules involving the orientational dependence of London's dispersion potential. The Maier-Saupe approach generates the pseudo-potential for one molecule interacting with an average field produced by the anisotropic molecular interactions. From this pseudo-potential the average thermodynamic and structural properties of interest for the nematic phase can be predicted. The simple theory gives qualitatively good agreement with certain experimental results [6, 7]. However, quantitatively their theory was found to overestimate many properties, including the transitional entropy, the nematic-isotropic transition temperature and the orientational order parameters [7]. Subsequently many simulation studies allowed the approximations and assumptions in the Maier-Saupe theory to be tested in depth [8-11]. This has lead to a whole body of research using the simple theory to improve our understanding of the liquid crystal phases not limited simply to nematics and nematic mixtures [12-14]. Thus, extensive investigations have been carried out and applied to many other liquid crystal phases such as smectic A [15, 16], chiral nematic [17] and biaxial nematics [18, 19]. The last member on this list of phases forms the focus of our investigation.

* Corresponding author: G.R.Luckhurst@soton.ac.uk

The biaxial nematic phase was first predicted by Freiser [18] who suggested, based on a molecular field theory, that non-uniaxial nematic phases might be possible. Since the initial prediction, there has been a huge interest in both theoretical and experimental studies of this novel liquid crystal phase as evidenced by recent comprehensive reviews [20, 21]. The seminal work of Freiser suggested, implicitly, that lath-like molecules with a $D_{2h}$ point group symmetry could self-organize and form a biaxial nematic phase, also with $D_{2h}$ symmetry. Subsequently, a theory also based on the molecular field approximation was developed by Straley to model the biaxial nematic assuming, explicitly, that the component molecules possessed $D_{2h}$ symmetry [19]. Following these seminal studies [18, 19], there were other theories [12, 22] and simulations [13, 23, 24] of the biaxial nematic, which assumed that this phase has $D_{2h}$ symmetry. However, recent NMR investigations of some nematic phases thought to be biaxial suggested that they could have a lower symmetry, namely $C_{2h}$ [25, 26]. Indeed, following the suggestion by Freiser [18], there has been a number of theories dedicated to a variety of nematic phases having a range of different point group symmetries, such as $C_{nh}$ and $D_{nh}$ (where n ≥ 2) [27-29]. A succinct discussion on the subject is given in reference [30]. However, none of these studies considered the application of the molecular field theory to the $C_{2h}$ biaxial nematic until Luckhurst *et al.*[30] developed their theory. In this study, the molecules that constituted the nematic phase were also taken to have $C_{2h}$ point group symmetry. A more specific model based on molecules formed from four Gay-Berne particles arranged to have $C_{2h}$ symmetry has been proposed by Gorkunov *et al.*[31]. This was found to form uniaxial and $D_{2h}$ biaxial nematic phases but not a $C_{2h}$ biaxial nematic. It would seem that the interaction parameters related to the deviation of the model molecules from orthorhombic to monoclinic shape proved to be too small for this phase to form. Subsequently computer simulations of the nematic phases formed by the simpler calamitic molecules [30] have also been carried out [32] in order to complement the theory.

Within both the molecular field theory and simulations the $C_2$ rotation axis of the calamitic molecules was taken to be parallel to the molecular long axis, as shown in Fig. 1. In constructing this idealised shape we start with an orthorhombic molecule with $z$ along the major axis, $y$ along the larger of the minor axes and $x$ along the smallest of these (see Fig.1). This hypothetical object is then divided into two halves along the $zy$ plane. These halves are moved with respect to each other along the $y$ axis to give the $C_{2h}$ shape shown in Fig. 1 together with the coordinate system.

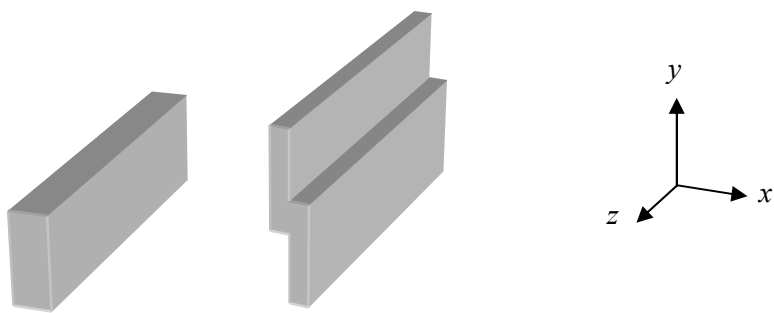

FIG. 1. The definition of the axes for the original orthorhombic molecule and hypothetical calamitic molecules having $D_{2h}$ and $C_{2h}$ point group symmetry [30], respectively.

With this location of the axes, a set of nine independent second rank order parameters was obtained, which reflected the symmetry of the three nematic phases and six intermolecular coupling parameters. In order to make the molecular field theory of the model more manageable, the those order parameters which vanish in the high order limit were eliminated as were the coupling parameters which are responsible for them, as suggested by Sonnet *et al.* [33] for molecules with $D_{2h}$ symmetry. One of the main features of this parameterization is that it preserves the interaction between the calamitic molecules. Thus the alignment of the molecular $z$-axis in the nematic phase is stronger compared to the alignment of the other two minor molecular axes in the nematic phase.

With this parameterization procedure, the simplified model of Luckhurst *et al*. [30] could be related to a molecule with a monoclinic shape and its long axis parallel to the $C_2$ rotational symmetry axis; we shall refer to this as Model 1. In order to extend this model to other monoclinic structures, we consider systems where the $C_2$ axis is along one or other of the two molecular short axes. To do this we shall build on Model 1 [30].

The layout of this paper is the following. In Sec. II we determine the coupling parameters for molecules with $C_{2h}$ symmetry, where the $C_2$ axis is defined along the *x* or *y* molecular axes. In addition the second rank orientational order parameters for these molecules in the nematic phases $NU$, $ND_{2h}$ and $NC_{2h}$ are considered. In Sec. III we construct the nematic Helmholtz free energy for our models based on the Gibbs entropy. Variational minimisation of the free energy with respect to the singlet orientational distribution using the analysis proposed by de Gennes [34] gives the potential of mean torque This method is more general than the Maier-Saupe approach, but the two results are similar[30, 35]. An extension of the Luckhurst-Romano [36] parameterization to the $C_{2h}$ model is proposed. In Sec. IV we describe the programs used to perform the calculations. Our results are given in Sec.V for the order parameters and the phase maps for the two new simplified models which we call Models 2 and 3. We compare the phase behaviour of these models with that of Model 1 [30] where the $C_2$ axis is parallel to the molecular long axis. Finally, our conclusions are summarised in Sec. VI.

**II. ORDER PARAMETERS AND COUPLING PARAMETERS**

We begin by defining the second rank order parameters needed to identify the three nematic phases formed from molecules having $C_{2h}$ symmetry with which we are concerned. These molecules are shown in Fig. 2 and have their $C_2$ rotation axis aligned orthogonal to the molecular long axis, *z*; in Model 2 this is along the long minor axis, *x*, and for Model 3 it is along the shortest axis, $y$. These hypothetical molecules were constructed from the same initial orthorhombic shape. For Model 2 this was cut in two halves along the $xy$ plane and then by moving one block with respect to the other along the $y$ axis. This gives the $C_{2h}$ molecule shown in Fig. 2(a) with the $C_2$ axis along $x$. For Model 3 the orthorhombic molecule was again cut into two halves but now one half is moved with respect to the other along the $x$ axis resulting in the $C_2$ axis being along $y$ as in Fig. 2(b). The orientational order parameters for these molecules are defined in terms of the averages of the Wigner functions $D^2_{pm}(\Omega)$ where $\Omega$ denotes the Euler angles giving the orientation of the molecular frame in the laboratory frame described by the director orientations. The subscripts $p$ and $m$ in the order parameters, $\langle D^2_{pm} \rangle$, relate to the properties of the phase and the molecules, respectively. The independent non-zero order parameters depend on the symmetry of the phase and of the molecules [37]. For our systems we have obtained a set of order parameters for each of the three distinct nematic phases, namely those with uniaxial, $D_{2h}$ biaxial and $C_{2h}$ biaxial symmetry for both Models 2 and 3. The two possible sets of order parameters for the two models and three phases are tabulated in Table 1. These combinations of Wigner function based order parameters are particularly convenient. Thus in the $NU$ nematic there are just three non-zero order parameters, on forming the $ND_{2h}$ phase three more order parameters are added, giving a total of six, and a further three appear for the $NC_{2h}$ nematic, giving a total of nine. For comparison the corresponding order parameters for Model 1 [30] are given in Table 2.

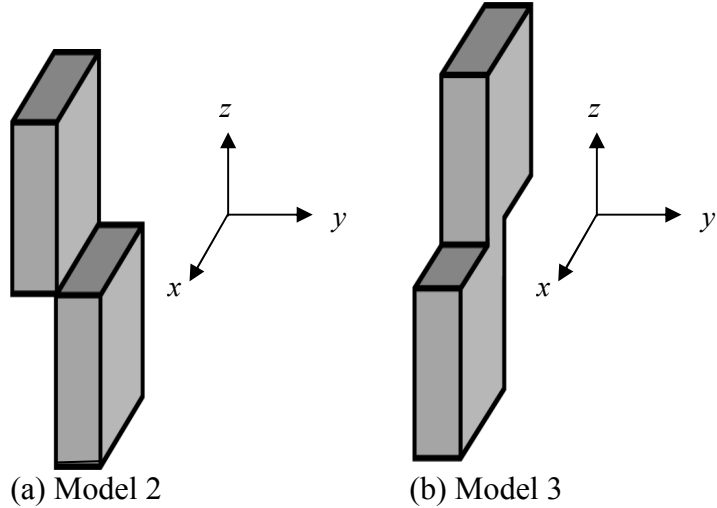

(a) Model 2  (b) Model 3

FIG.2. The two models proposed for the monoclinic molecule with the $C_2$ axis orthogonal to the molecular long axis, $z$. (a) Model 2 is where the $C_2$ symmetry axis is defined as the long minor axis, $x$, and (b) Model 3 is where the $C_2$ symmetry axis is defined as the short minor axis, $y$.

| Phase | Second Rank Order Parameters for Model 2 | Second Rank Order Parameters for Model 3 |
|---|---|---|
| $NU$ | $\langle R_{00} \rangle = \langle D_{00}^2 \rangle$ | $\langle R_{00} \rangle = \langle D_{00}^2 \rangle$ |
| | $\langle I_{01} \rangle = \left( \langle D_{01}^2 \rangle + \langle D_{0-1}^2 \rangle \right)/2i$ | $\langle R_{01} \rangle = \left( \langle D_{01}^2 \rangle - \langle D_{0-1}^2 \rangle \right)/2$. |
| | $\langle R_{02} \rangle = \left( \langle D_{02}^2 \rangle + \langle D_{0-2}^2 \rangle \right)/2$ | $\langle R_{02} \rangle = \left( \langle D_{02}^2 \rangle + \langle D_{0-2}^2 \rangle \right)/2$ |
| $ND_{2h}$ | $\langle R_{20} \rangle = \left( \langle D_{20}^2 \rangle - \langle D_{-20}^2 \rangle \right)/2$ | $\langle R_{20} \rangle = \left( \langle D_{20}^2 \rangle - \langle D_{-20}^2 \rangle \right)/2$ |
| | $\langle I_{21} \rangle = \left( \langle D_{21}^2 \rangle + \langle D_{-2-1}^2 \rangle + \langle D_{-21}^2 \rangle + \langle D_{2-1}^2 \rangle \right)/2i$ | $\langle R_{21} \rangle = \left( \langle D_{21}^2 \rangle - \langle D_{-2-1}^2 \rangle + \langle D_{-21}^2 \rangle - \langle D_{2-1}^2 \rangle \right)/2$ |
| | $\langle R_{22} \rangle = \left( \langle D_{22}^2 \rangle + \langle D_{-2-2}^2 \rangle + \langle D_{-22}^2 \rangle + \langle D_{2-2}^2 \rangle \right)/2$ | $\langle R_{22} \rangle = \left( \langle D_{22}^2 \rangle + \langle D_{-2-2}^2 \rangle + \langle D_{-22}^2 \rangle + \langle D_{2-2}^2 \rangle \right)/2$ |
| $NC_{2h}$ | $\langle I_{10} \rangle = \left( \langle D_{10}^2 \rangle + \langle D_{-10}^2 \rangle \right)/2i$ | $\langle R_{10} \rangle = \left( \langle D_{10}^2 \rangle - \langle D_{-10}^2 \rangle \right)/2$ |
| | $\langle R_{11}^s \rangle = \left( \langle D_{11}^2 \rangle + \langle D_{-1-1}^2 \rangle + \langle D_{-11}^2 \rangle + \langle D_{1-1}^2 \rangle \right)/2$ | $\langle R_{11}^a \rangle = \left( \langle D_{11}^2 \rangle + \langle D_{-1-1}^2 \rangle - \langle D_{-11}^2 \rangle - \langle D_{1-1}^2 \rangle \right)/2$ |
| | $\langle I_{12} \rangle = \left( \langle D_{12}^2 \rangle + \langle D_{-1-2}^2 \rangle + \langle D_{-12}^2 \rangle + \langle D_{1-2}^2 \rangle \right)/2i$. | $\langle R_{12} \rangle = \left( \langle D_{12}^2 \rangle - \langle D_{-1-2}^2 \rangle + \langle D_{1-2}^2 \rangle - \langle D_{-12}^2 \rangle \right)/2$ |

Table 1. Definition of the non-zero second rank orientational order parameters for Models 2 and 3, according to the phase symmetry.

| Phase | Second Rank Order Parameters |
|---|---|
| $NU$ | $\langle R_{00} \rangle = \langle D_{00}^2 \rangle$ |
| | $\langle R_{02} \rangle = \left( \langle D_{02}^2 \rangle + \langle D_{0-2}^2 \rangle \right)/2$ |
| | $\langle I_{02} \rangle = \left( \langle D_{02}^2 \rangle - \langle D_{0-2}^2 \rangle \right)/2$ |
| $ND_{2h}$ | $\langle I_{20} \rangle = \left( \langle D_{20}^2 \rangle - \langle D_{-20}^2 \rangle \right)/2i$ |
| | $\langle R_{22}^a \rangle = \left[ \left( \langle D_{22}^2 \rangle + \langle D_{-2-2}^2 \rangle \right) - \left( \langle D_{-22}^2 \rangle + \langle D_{2-2}^2 \rangle \right) \right]/2$ |
| | $\langle I_{22}^a \rangle = \left[ \left( \langle D_{22}^2 \rangle - \langle D_{-2-2}^2 \rangle \right) - \left( \langle D_{-22}^2 \rangle - \langle D_{2-2}^2 \rangle \right) \right]/2i$ |
| $NC_{2h}$ | $\langle I_{20} \rangle = \left( \langle D_{20}^2 \rangle - \langle D_{-20}^2 \rangle \right)/2i$ |
| | $\langle R_{22}^a \rangle = \left[ \left( \langle D_{22}^2 \rangle + \langle D_{-2-2}^2 \rangle \right) - \left( \langle D_{-22}^2 \rangle + \langle D_{2-2}^2 \rangle \right) \right]/2$ |
| | $\langle I_{22}^a \rangle = \left[ \left( \langle D_{22}^2 \rangle - \langle D_{-2-2}^2 \rangle \right) - \left( \langle D_{-22}^2 \rangle - \langle D_{2-2}^2 \rangle \right) \right]/2i$ |

Table 2. Definition of the non-zero second rank orientational order parameters for Model 1 taken from the previous study by Luckhurst *et al*. [30], and grouped according to the phase symmetry .

According to the molecular field theory proposed by de Gennes the internal energy is given by products of the dominant order parameters [34]; since he was only concerned with uniaxial molecules in a uniaxial nematic phase this was fairly simple. However, for molecules and phases of lower symmetry the situation is more complex. For these we shall write the internal energy as a sum of invariants formed from the order parameters, $\langle D^2_{pm} \rangle$, as [35]

$$\langle U \rangle = -\frac{1}{2} \sum_{mnp} u_{2mn} \langle D^2_{pm} \rangle \langle D^2_{-pn} \rangle. \tag{1}$$

Here the proportionality or coupling parameters have the same form as the intermolecular interaction coefficients, used in the S-function based expansion of the pair potential [38], to which they are related.

The form of the coupling parameters is determined by the molecular symmetry [38]. In Model 1, where the molecular long axis is parallel to the $C_2$ axis and labelled the $z$-axis [30], the symmetry elements of the molecules include inversion $i$, the two-fold rotation along the $z$-axis $C_2^{(z)}$ and an $xy$-reflection plane, $\sigma_{xy}$. Using the symmetry arguments given by Stone [38] $m$ and $n$ in Eq.(1) can only have values of either 0 or ±2. On the other hand, when the $C_2$ axis is orthogonal to the molecular z-axis as shown in Fig. 2 for Models 2 and 3, $m$ and $n$ can have values of 0, ±1 and ±2.

In Model 2 (see Fig. 2(a)), the symmetry elements for a molecule are $i$, $C_2^{(x)}$ and $\sigma_{yz}$, which give the following relations between the coupling parameters:

$$\begin{aligned}
u_{210} &= u_{2-10} = u_{201} = u_{20-1}, \\
u_{211} &= u_{2-11} = u_{21-1} = u_{2-1-1}, \\
u_{221} &= u_{2-21} = u_{22-1} = u_{2-2-1} = u_{212} = u_{21-2} = u_{2-12} = u_{2-1-2}, \\
u_{220} &= u_{2-20} = u_{202} = u_{20-2}, \\
u_{222} &= u_{2-22} = u_{22-2} = u_{2-2-2}.
\end{aligned} \tag{2}$$

There are then six independent coupling parameters for Model 2 namely $u_{200}$, $u_{210}$, $u_{220}$, $u_{221}$, $u_{211}$ and $u_{222}$. The total number of independent intermolecular coefficients is, necessarily, the same as that in the previous model [30]. Similarly for Model 3 the symmetry elements are $i$, $C_2^{(y)}$ and $\sigma_{xz}$, resulting in slightly different relationships between the coupling parameters, namely

$$u_{210} = -u_{2-10} = u_{201} = -u_{20-1},$$
$$u_{211} = -u_{2-11} = -u_{21-1} = u_{2-1-1},$$
$$u_{221} = u_{2-21} = -u_{22-1} = -u_{2-2-1} = u_{212} = u_{21-2} = -u_{2-12} = -u_{2-2-1}, \quad (3)$$
$$u_{220} = u_{2-20} = u_{202} = u_{20-2},$$
$$u_{222} = u_{2-22} = u_{22-2} = u_{2-2-2}.$$

Even though the independent coupling parameters are the same for both Models 2 and 3, the relationships (see Eqs.(2) and (3)) between them differ in some signs begin to emerge when considering their reality [30]. To ensure the internal energy is real, that is the condition $\langle U^* \rangle = \langle U \rangle$ must be satisfied, we require

$$u_{2mn}{}^* = (-1)^{m+n} u_{2-m-n}. \quad (4)$$

In the case of molecules with $D_{2h}$ symmetry [30], all of the independent coefficients are real. The situation for the $C_{2h}$ Model 1 [30], where the $C_2$ axis is parallel to the long axis, some of the intermolecular coefficients are complex. In Model 2, $u_{210}$ and $u_{221}$ are purely imaginary while the others are real,

$$u_{210}{}^* = -u_{210},$$
$$u_{221}{}^* = -u_{221}. \quad (5)$$

Interestingly, we find that for Model 3, all of the coefficients are real because the set of order parameters defined according to the combination of the Wigner functions for Model 3 are also real (see Table 1).

### III. MOLECULAR FIELD THEORY

Armed with the order parameters and coupling parameters for Models 2 and 3 of molecules having $C_{2h}$ symmetry we now construct the molecular field theories for them by applying a methodology analogous to that used by Luckhurst *et al.* [30]. Initially, the averaged anisotropic internal energy $\langle U \rangle$ of the system is defined using the orientational order parameters written in terms of Wigner rotation matrices $\langle D_{mn}^2 \rangle$,

$$\langle U \rangle = -\frac{1}{2} \sum_{mnp} u_{2mn} \langle D_{pm}^2 \rangle \langle D_{-pn}^2 \rangle. \quad (6)$$

Here,

$$\langle D_{mn}^2 \rangle = \int D_{mn}^2(\Omega) f(\Omega) d\Omega ,$$

(7)

where $f(\Omega)$ is the singlet orientational distribution function. The entropy of the system is given by the Gibbs entropy formula

$$S = -k_B \int f(\Omega) \ln f(\Omega) d\Omega \qquad (8)$$

and $k_B$ is the Boltzmann constant. Combining the results for the internal energy and entropy we can construct the Helmholtz free energy, $A$, of the system as

$$A = -\frac{1}{2} \sum_{mnp} u_{2mn} \langle D_{pm}^2 \rangle \langle D_{-pn}^2 \rangle + k_B T \int f(\Omega) \ln f(\Omega) d\Omega , \qquad (9)$$

where $T$ is the temperature. The unknown singlet orientational distribution function is now determined by minimizing the free energy with respect to $f(\Omega)$. This is subject to the constraints that the distribution function is normalised and that the order parameters are related to the distribution function by Eq. (7); this process gives the potential of mean torque as

$$U(\Omega) = -\frac{1}{2} \sum_{mnp} u_{2mn} \langle D_{pm}^2 \rangle D_{-pn}^2(\Omega) . \qquad (10)$$

Using the properties of the coupling parameters and orientational order parameters derived in Sec. II for the two models the potential of mean torque for each model is found to contain three terms identified as the driving force for the formation of the uniaxial ($NU$), the $D_{2h}$ biaxial ($ND_{2h}$) and the $C_{2h}$ biaxial ($NC_{2h}$) nematics, namely $U_U(\Omega)$, $U_{D_{2h}}(\Omega)$ and $U_{C_{2h}}(\Omega)$,

$$U(\Omega) = U_U(\Omega) + U_{D_{2h}}(\Omega) + U_{C_{2h}}(\Omega) . \qquad (11)$$

The components of the potential of mean torque for Model 2 are

$$U_U(\Omega) = -\big[ \big( \langle R_{00} \rangle + 2\lambda_{10} \langle R_{01} \rangle + 2\lambda_{02} \langle R_{02} \rangle \big) R_{00}(\Omega)$$
$$+ \big( 2\lambda_{10} \langle R_{00} \rangle + 4\lambda_{21} \langle R_{02} \rangle + 4\lambda_{11} \langle I_{01} \rangle \big) I_{01}(\Omega) \qquad (12)$$
$$+ \big( 2\lambda_{20} \langle R_{00} \rangle + 4\lambda_{12} \langle I_{01} \rangle + 4\lambda_{22} \langle R_{02} \rangle \big) R_{02}(\Omega) \big],$$

$$U_{D_{2h}}(\Omega) = -2\Big[\big(\langle R_{20}\rangle + \lambda_{10}\langle I_{21}\rangle + \lambda_{02}\langle R_{22}\rangle\big)R_{20}(\Omega)$$
$$+\big(\lambda_{10}\langle R_{20}\rangle + \lambda_{21}\langle R_{22}\rangle + \lambda_{11}\langle I_{21}\rangle\big)I_{21}(\Omega) \quad (13)$$
$$+\big(\lambda_{20}\langle R_{20}\rangle + \lambda_{12}\langle I_{21}\rangle + \lambda_{22}\langle R_{22}\rangle\big)R_{22}(\Omega)\Big],$$

$$U_{C_{2h}}(\Omega) = -2\Big[\big(-\langle I_{10}\rangle + \lambda_{10}\langle R_{11}^s\rangle - \lambda_{02}\langle I_{12}\rangle\big)I_{10}(\Omega)$$
$$+\big(\lambda_{10}\langle I_{10}\rangle + \lambda_{21}\langle I_{12}\rangle + \lambda_{11}\langle R_{11}^s\rangle\big)R_{11}^s(\Omega) \quad (14)$$
$$+\big(-\lambda_{20}\langle I_{10}\rangle + \lambda_{12}\langle R_{11}^s\rangle - \lambda_{22}\langle I_{12}\rangle\big)I_{12}(\Omega)\Big].$$

The potential of mean torque is scaled with the coupling parameter $u_{200}$ and so we define the scaled coupling factors as

$$\lambda_{mn} = |u_{2mn}|/u_{200}. \quad (15)$$

For Model 3, the potential of mean torque is

$$U_U(\Omega) = -\Big[\big(\langle R_{00}\rangle - 2\lambda_{10}\langle R_{01}\rangle + 2\lambda_{02}\langle R_{02}\rangle\big)R_{00}(\Omega)$$
$$+\big(2\lambda_{10}\langle R_{00}\rangle + 4\lambda_{21}\langle R_{02}\rangle + 4\lambda_{11}\langle R_{01}\rangle\big)R_{01}(\Omega) \quad (16)$$
$$+\big(2\lambda_{20}\langle R_{00}\rangle + 4\lambda_{12}\langle R_{01}\rangle + 4\lambda_{22}\langle R_{02}\rangle\big)R_{02}(\Omega)\Big],$$

$$U_{D_{2h}}(\Omega) = -2\Big[\big(\langle R_{20}\rangle + \lambda_{10}\langle R_{21}\rangle + \lambda_{02}\langle R_{22}\rangle\big)R_{20}(\Omega)$$
$$+\big(\lambda_{10}\langle R_{20}\rangle + \lambda_{21}\langle R_{22}\rangle + \lambda_{11}\langle R_{21}\rangle\big)R_{21}(\Omega) \quad (17)$$
$$+\big(\lambda_{20}\langle R_{20}\rangle + \lambda_{12}\langle R_{21}\rangle + \lambda_{22}\langle R_{22}\rangle\big)R_{22}(\Omega)\Big],$$

$$U_{C_{2h}}(\Omega) = -2\Big[\big(-\langle R_{10}\rangle + \lambda_{10}\langle R_{11}^a\rangle + \lambda_{02}\langle R_{12}\rangle\big)R_{10}(\Omega)$$
$$+\big(\lambda_{10}\langle R_{10}\rangle + \lambda_{21}\langle R_{12}\rangle + \lambda_{11}\langle R_{11}^a\rangle\big)R_{11}^a(\Omega) \quad (18)$$
$$+\big(\lambda_{20}\langle R_{10}\rangle + \lambda_{12}\langle R_{11}^a\rangle + \lambda_{22}\langle R_{12}\rangle\big)R_{12}(\Omega)\Big].$$

In both models the potential of mean torque contains nine order parameters and five scaled coupling parameters. Given a set of scaled coupling parameters with positive values, we can solve the molecular field theory by minimizing the Helmholtz free energy [30, 33] with respect to the order parameters and obtain the phase behaviour of our system via the temperature dependence of the order parameters. However, this parameter space is very large, indeed each of these models involves nine multi-dimensional minimizations and so the prediction of the phase behaviour is a formidable task. To reduce the complexity of such a problem, but at the same time, still to be able to capture the essential physics of the system, following Sonnet *et al.* [33] and Luckhurst *et al.* [30], we assume that those order parameters which vanish in the high order limit may be set to zero. The order parameters that do vanish are those for which $p \neq m$. To ensure that they remain zero the associated scaled coupling parameter $\lambda_{mn}$, linking the order parameter involving $m$ with the angular function containing $n$, is set to zero, again when $m \neq n$. Those remaining terms, which involve $p \neq m$, belong to the dominant order parameters. The remaining order parameters for Model 2 after this reduction are $\langle R_{00} \rangle$, $\langle R_{11}^s \rangle$ and $\langle R_{22} \rangle$, while for Model 3 they are $\langle R_{00} \rangle$, $\langle R_{11}^a \rangle$ and $\langle R_{22} \rangle$. For Model 1 [30] the surviving order parameters are $\langle R_{00} \rangle$, $\langle R_{22} \rangle \equiv \langle R_{22}^s \rangle$ and $\langle R_{22}^a \rangle$. For Model 2 in the $NC_{2h}$ phase all three order parameters $\langle R_{00} \rangle$, $\langle R_{11}^a \rangle$ and $\langle R_{22} \rangle$ are non-zero, for the $ND_{2h}$ phase the two order parameters $\langle R_{00} \rangle$ and $\langle R_{22} \rangle$ are non-zero and for the $NU$ phase only $\langle R_{00} \rangle$ survives. The truncated pseudo-potentials for Models 2 and 3 are

$$U_{\text{trun}}^2(\Omega) = -\left[ \langle R_{00} \rangle R_{00}(\Omega) + \lambda_{22} \langle R_{22} \rangle R_{22}(\Omega) + 2\lambda_{11} \langle R_{11}^s \rangle R_{11}^s(\Omega) \right], \tag{19}$$

$$U_{\text{trun}}^3(\Omega) = -\left[ \langle R_{00} \rangle R_{00}(\Omega) + \lambda_{22} \langle R_{22} \rangle R_{22}(\Omega) + 2\lambda_{11} \langle R_{11}^a \rangle R_{11}^a(\Omega) \right], \tag{20}$$

respectively. With these potentials, we write their Helmholtz free energies as

$$A^i / u_{200} = (1/2)\left[ \langle R_{00} \rangle^2 + \lambda_{22} \langle R_{22} \rangle^2 + 2\lambda_{11} \langle R_{11}^i \rangle^2 \right] - T^* \ln Q, \tag{21}$$

where $i = s$ or $a$ according to the particular model (see, Tables 1 and 2), $T^*$ ($\equiv k_B T / u_{200}$) is the scaled temperature and $Q$ is the orientational partition function

$$Q = \int \exp\left(-U_{\text{trun}}^j(\Omega) / T^*\right) d\Omega, \tag{22}$$

Here $j = 2$ or 3 depending on which system we are considering (see Eqs. (19) and (20), respectively). With this parameterization procedure, the behaviour of the system can be calculated more easily and depends on only the two scaled coupling parameters $\lambda_{11}$ and $\lambda_{22}$.

## IV. COMPTUTAIONAL DETAILS

The numerical minimization of the Helmholtz free energy was performed using a FORTRAN program, compiled by the gfortran compiler in LINUX, which also incorporates subroutines provided by Numerical Recipes [39]. The algorithm used in the minimization program is the Broyden-Fletcher-Goldfarb-Shanno (BFGS) algorithm. This applies to the quasi-Newton method, where the approximated Hessian was used in calculating the direction, on the free energy surface, along which the free energy would decrease to its local minimum. In addition, the numerical integration employed in calculating the partition function and the first derivatives of the free energy with respect to the three order parameters, needed to ensure that a minimum has been reached, is based on the Gaussian quadrature method [39].

## V. RESULTS AND DISCUSSION

The first step in our study of the nematic phases formed by Models 2 and 3 was the selection of the scaled coupling parameters $\lambda_{11}$ and $\lambda_{22}$ in the respective potentials of mean torque. We were guided in our choice by the values of the analogous parameters $\lambda_s$ and $\lambda_a$ used in the calculations for Model 1 [30]. Here $\lambda_s$ is equivalent to $\lambda_{22}$ and drives the formation of the $ND_{2h}$ phase; in the original calculations it was given the values 0.2, 0.3 and 0.4, accordingly we shall use the same values for the new models. The second parameter $\lambda_a$ drives the formation of the $NC_{2h}$ phase and was allowed to vary continuously over the range from 0 to 0.6; this was found to reveal a rich phase behaviour [30]. Its analogue for the new models is the scaled coupling parameter, $\lambda_{11}$, which has been studied over the same range. However, we should note that the values of the coupling parameters for the idealised molecules in the three models are certainly different. Unfortunately it is difficult, even knowing the dimensions of these simple objects (see Figs. 1 and 2) to estimate the coupling supertensor, $u_{2mn}$, indeed, based on the excluded volumes, this must be undertaken numerically [40]. None the less it might be expected that for similar and small displacements of the blocks used to construct the $C_{2h}$ molecules that the coupling tensor components $u_{200}$ and $u_{2mn}$ will be similar for the three model molecules. It is not so easy to assess the similarity for the tensor components that drive the formation of the $NC_{2h}$ phase for these three models but given the manipulation of the half blocks they might well be expected to be similar.

For the particular ranges of the coupling parameters $\lambda_{11}$ and $\lambda_{22}$, we have minimized, numerically, the Helmholtz free energy in Eq. (21) at a selection of scaled temperatures by varying the order parameters $\langle R_{00} \rangle$, $\langle R_{11}^s \rangle$ and $\langle R_{22} \rangle$. The most stable state was achieved for the set of order parameters giving a global minimum of the free energy. These calculations gave the temperature dependence of the orientational order parameters for particular choices of $\lambda_{11}$ and $\lambda_{22}$. The scaled temperature at which the order parameter defining the lower temperature phase passes from zero to non-zero is identified as the transition temperature. The order of this transition is determined by whether the change in the defining order parameter is discontinuous (first order) or continuous (second order) within the computational error.

From our results it was found that, for the same values of $\lambda_{11}$ and $\lambda_{22}$, the phase behaviour of Models 2 and 3 are quantitatively the same. This is to be expected since both models have the $C_2$ axis orthogonal to the molecular long axis. This equivalence is shown analytically using the molecular field theory approach in the Appendix. In view of this we shall only consider the results for Model 2. Never the less, at a practical level even for the same values of $\lambda_{11}$ and $\lambda_{22}$, the shapes of the constituent molecules for the two models are expected to differ. However, it is not our major concern in this paper to take account of how the molecular shape changes but to explore how the phase behaviour changes with the orientation of the $C_2$ axis with respect to the molecular long axis.

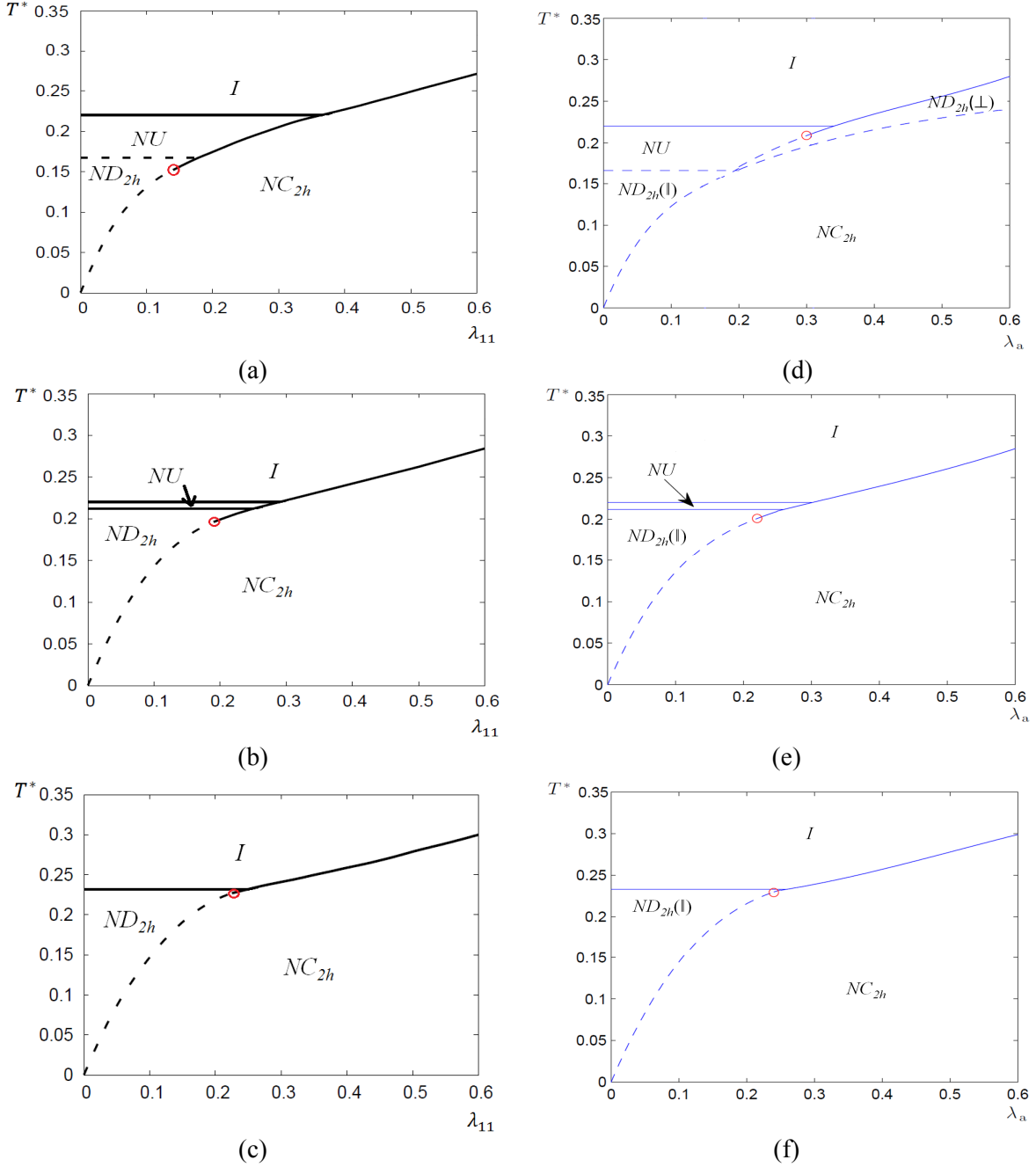

FIG. 3. The phase maps calculated from the truncated potential of mean torque for Model 2. These show the transition temperatures and phase sequences when the scaled coupling coefficient, $\lambda_{11}$, varies from 0 to 0.6, while $\lambda_{22}$ is fixed at (a) 0.2, (b) 0.3 and (c) 0.4. In addition, (d), (e) and (f) show the phase maps for Model 1 when $\lambda_s$ is equal to 0.2, 0.3 and 0.4, respectively (adapted with permission from ref.[30]). The dashed lines indicate second order phase transitions and the solid lines show first order phase transitions; the red circles denote the associated tricritical points.

The phase maps show how the transition temperatures between the phases change with the variation in one of the scaled coupling parameters occurring in the potential of mean torque; they are given in Fig. 3. Those for Model 2 appear in (a), (b) and (c) where $\lambda_{11}$ is varied continuously while $\lambda_{22}$ is held fixed at 0.2, 0.3 and 0.4 for the plots in Fig. 3(a), (b) and (c), respectively. We start our discussion of these maps with that in Fig. 3(a) where $\lambda_{22}$ is 0.2. For $\lambda_{11}$ of zero the system shows the phase sequence $ND_{2h} - NU - I$ where the transition temperatures are in agreement with those obtained previously [30, 33]. These do not change as $\lambda_{11}$ increases since the truncated versions of the potentials of mean torque, $U_U(\Omega)$ and $U_{D_{2h}}(\Omega)$, do not depend on this coupling parameter (see Eq. (19)). However, as $\lambda_{11}$ grows so the $NC_{2h}$ phase is introduced into the map and increases in extent so that the phase sequences $NC_{2h} - ND_{2h} - NU - I$, $NC_{2h} - NU - I$ and $NC_{2h} - I$ are observed. In addition, the $NU - I$ transition is first order whereas the $ND_{2h} - NU$ transition is second order, as expected [30, 33]. In contrast the $NC_{2h} - ND_{2h}$ transition is second order for $\lambda_{11}$ less than about 0.14 and then becomes first order as is the $NC_{2h} - I$ transition.

The behaviour found for Model 2 with this particular parameterisation contrasts dramatically with that found previously for Model 1 having a comparable choice of coupling parameters [30]. The phase map for this is shown in Fig. 3(d) and reveals a new biaxial nematic labelled $ND_{2h}(\perp)$ not found for Model 2. An idealised structure of this new phase is sketched in Fig.

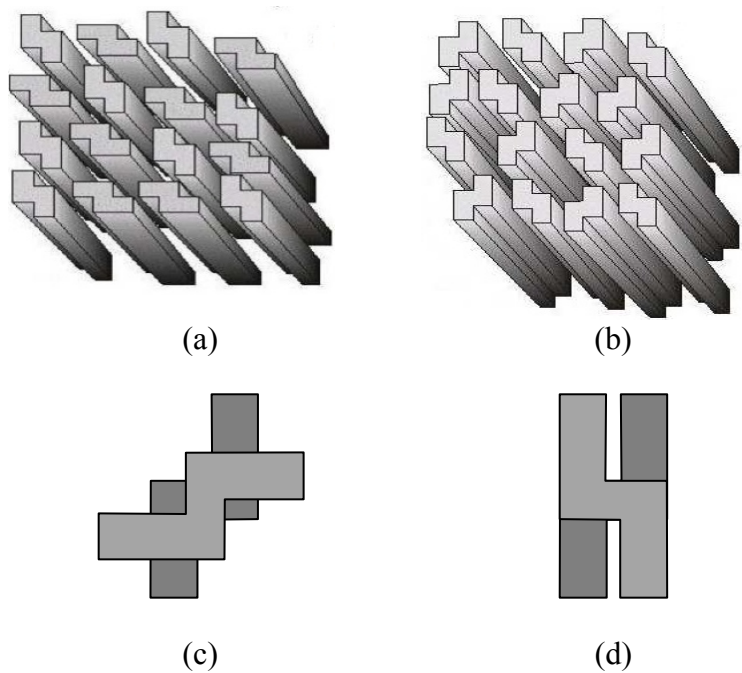

(a)   (b)

(c)   (d)

FIG. 4. The idealised organisation of the molecules for Model 1 in (a) the $ND_{2h}(\perp)$ phase and (b) the $ND_{2h}(\parallel)$ phase. (Adapted with permission from [30]). The associated composite structures are formed by combining two molecules and shown in cross section, orthogonal to their $C_2$ axis, in (c) for $ND_{2h}(\perp)$ and (d) for $ND_{2h}(\parallel)$.

4(a) where it is seen that the equivalent short axes of neighbouring molecules are orthogonal. In addition, the long $C_2$ axes are antiparallel which is required because the phase has $D_{2h}$ point group symmetry. For comparison we also show the idealised structure of the $ND_{2h}(\parallel)$ phase in Fig. 4(b). Again the $C_2$ axes of half the molecules are antiparallel to those of the other half but now the minor axes are parallel. The key stabilising feature for these two $ND_{2h}$ phases is that the molecular long axes are either parallel or anti-parallel. Another insight into the structure of these two phases can be obtained from the composite structure formed by merging two molecules having their $C_2$ axes antiparallel [28, 30]. The orientation of the short axis of one molecule with respect to the other is rotated, internally [29], by 180º to give the structure shown as the cross section perpendicular to the $C_2$ axis in Fig. 4(d) for the $ND_{2h}(\parallel)$ phase and when the internal rotation is 90º the composite structure, given in Fig. 4(c), also has $D_{2h}$ symmetry for the $ND_{2h}(\perp)$ phase.

We can use the same approach to see why it is not possible for Model 2 to form the $ND_{2h}(\perp)$ phase. Since this phase has $D_{2h}$ point group symmetry, then in the idealised structure the $C_2$ axes of half the molecules must be antiparallel to those of the other half. In such a three dimensional structure, shown in Fig. 5(a), where a short and a long axis can both be parallel the structure is stable, giving what is, in effect, the $ND_{2h}(\parallel)$ phase. However, for the case when the short axes are perpendicular the long axes must also be perpendicular, as we can see in Fig. 5(b), and it is this orthogonality which destabilises the $ND_{2h}(\perp)$ phase. The composite structures formed from pairs of molecules with their $C_2$ axes antiparallel also provide a simple image to understand the stability as well as the symmetry. They are shown in Fig. 5(c) where one short and the long axes are parallel and in Fig. 5(d) where the same axes are now orthogonal; both structures are seen to have structures with $D_{2h}$ point group symmetry.

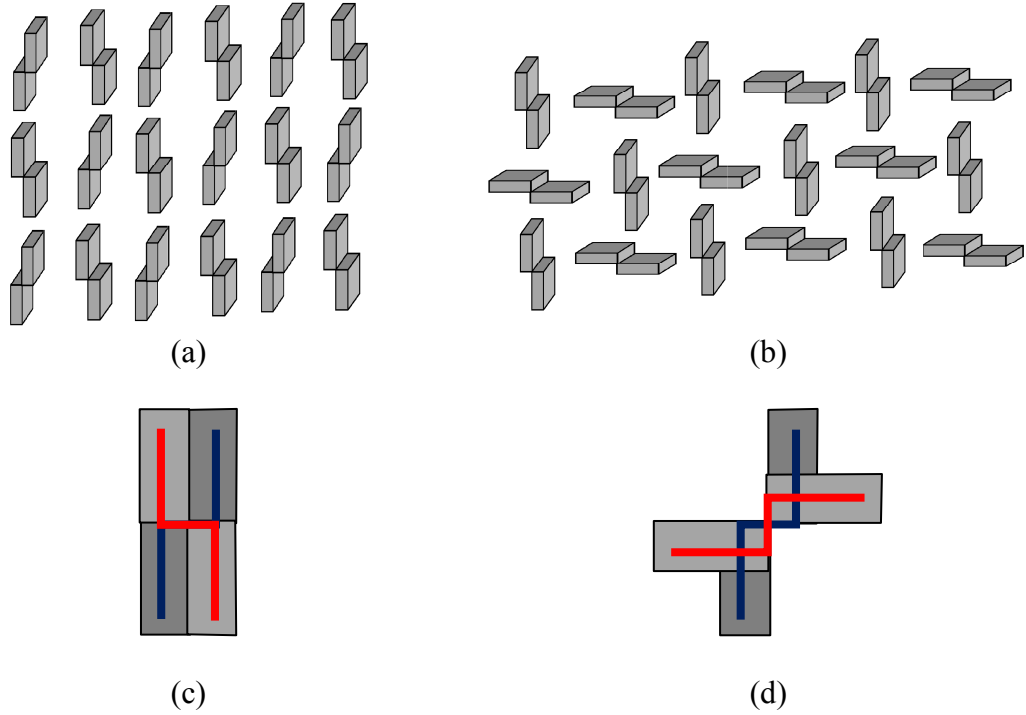

FIG. 5. The idealised organisation of the molecules for Model 2 in (a) the $ND_{2h}(\parallel)$ phase and (b) the $ND_{2h}(\perp)$ phase. The composite structures, associated with the phase structures, are formed by combining two molecules and shown in cross section, orthogonal to their $C_2$ axis, (c) for $ND_{2h}(\parallel)$. and (d) for $ND_{2h}(\perp)$

However, it is also apparent from these two composite structures that their anisotropies are significantly different. This is large for that in Fig. 5(c) leading to the stability of the $ND_{2h}(\parallel)$ phase but for that in Fig.5(d) it is small leading to the instability of the $ND_{2h}(\perp)$ phase. As a consequence we believe that, unlike Model 1, the $ND_{2h}(\perp)$ structure is not observed by Model 2 as long as it remains calamitic; that is the $C_2$ axis is orthogonal to the molecular long axis..

In the next set of calculations for Model 2 the biaxiality coupling parameter, $\lambda_{22}$, was increased to 0.3 and the phase map for this system is shown in Fig. 3(b). The phases formed are the same as those seen in Fig. 3(a) but, as expected, the stability of the $ND_{2h}$ phase has increased at the expense of the uniaxial nematic; indeed there is just a narrow band of $NU$ remaining. Both the $NU - I$ and the $ND_{2h} - NU$ phase transitions are now predicted to be first order, in keeping with previous studies [30, 33]. Transitions from the $NC_{2h}$ to the $ND_{2h}$ phase are found to exhibit tricritical behaviour but after the tricritical point at $\lambda_{11}$ of about 0.19 the transitions from the $NC_{2h}$ to the $ND_{2h}$ phase and then to the isotropic phase are first order. For comparison the phase map for Model 1 calculated with the same value of $\lambda_{22}$ and for the analogous parameter $\lambda_a$ which drives the formation of the $NC_{2h}$ phase is shown in Fig. 3(d). It is clear that there is a strong similarity between these phase maps for the two models and we shall return to this similarity later.

The final set of calculations was performed for Model 2 with $\lambda_{22}$ set equal to 0.4; the resultant phase map is given in Fig. 3(c). It is seen that for this large value of the scaled biaxial coupling parameter the $NU$ phase has been replaced entirely by the $ND_{2h}$ phase so that there are just two phase sequences $NC_{2h} - ND_{2h} - I$ and $NC_{2h} - I$. The $NC_{2h} - ND_{2h}$ phase transition is second order except just before the onset of the $NC_{2h} - I$ transition which is first order. The phase behaviour of Model 2 turns out to be essentially equivalent to that observed for Model 1 over a comparable range of the coupling parameter driving the formation of the $NC_{2h}$ phase.

To see just how similar such phase maps are we have studied the temperature dependence of the order parameters for Model 2 and Model 1 for the same value of $\lambda_{22}\,(\equiv \lambda_s)$ of 0.3 and for $\lambda_{11}\,(\equiv \lambda_a)$ equal to 0.15 (see Fig. 6) and $\lambda_{11}\,(\equiv \lambda_a)$ equal to 0.4 (see Fig. 7). Fig. 6 shows the phase sequence $NC_{2h} - ND_{2h} - NU - I$. The results for the two order parameters, $\langle R_{00} \rangle$ and $\langle R_{22} \rangle$, are essentially the same and indeed this has to be the case since the potentials of mean torque are identical. However, in the $NC_{2h}$ phase the defining order parameters $\langle R_{11}^s \rangle$ and $\langle R_{22}^a \rangle$ are slightly different causing the shift in the $NC_{2h} - ND_{2h}$ transition temperature to a slightly higher value for Model 2. On the other hand, parameterization with a higher value of $\lambda_{11}\,(\equiv \lambda_a)$, i.e. equal to 0.4 (see Fig. 7), gives the simple phase sequence $NC_{2h} - I$. The order parameters for this parameterization are presented separately according to the different order parameter definitions so as to make the comparison clearer, since the values of these order parameters come close to each other for this parameterization. Fig. 7 shows that Model 2 has, again, a higher $NC_{2h} - I$ transition temperature.

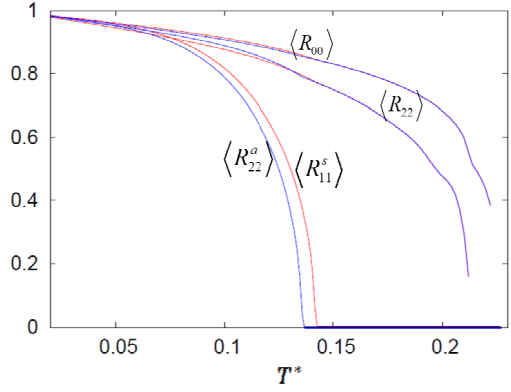

FIG. 6. The scaled temperature dependence of the orientational order parameters for Model 1 ($\langle R_{00}\rangle$, $\langle R_{22}\rangle$ and $\langle R_{22}^a\rangle$) and Model 2 ($\langle R_{00}\rangle$, $\langle R_{22}\rangle$ and $\langle R_{11}^s\rangle$) for $\lambda_{22}(\equiv\lambda_s) = 0.3$ and for $\lambda_{11}(\equiv\lambda_a) = 0.15$. The blue lines are given for Model 1, while red is for Model 2.

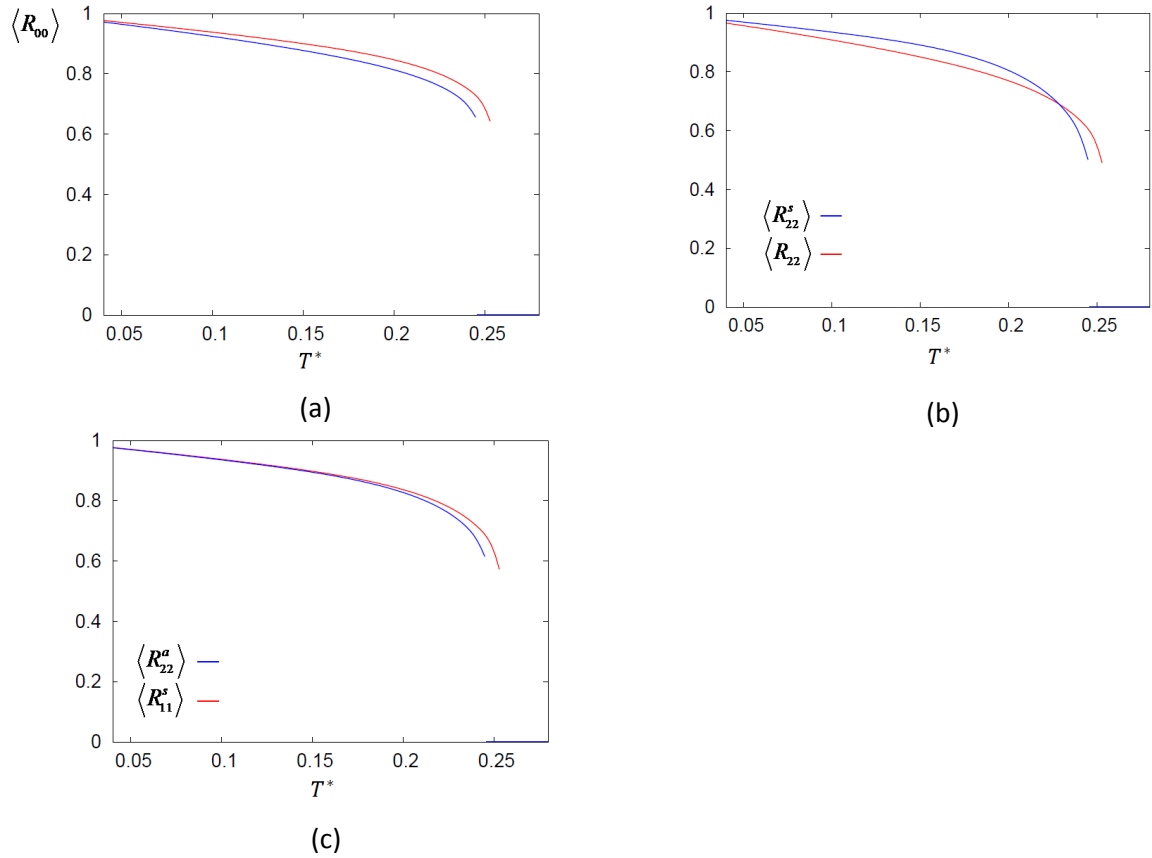

FIG. 7. The scaled temperature dependence of the orientational order parameters for Model 1 ($\langle R_{00}\rangle, \langle R_{22}^s\rangle$ and $\langle R_{22}^a\rangle$) and Model 2 ($\langle R_{00}\rangle, \langle R_{22}\rangle$ and $\langle R_{11}^s\rangle$) for $\lambda_{22}(\equiv\lambda_s) = 0.3$ and for $\lambda_{11}(\equiv\lambda_a) = 0.4$. The blue lines are given for Model 1, while red are given for Model 2. (a) $\langle R_{00}\rangle$ versus $T^*$. (b) $\langle R_{22}^s\rangle$ and $\langle R_{22}\rangle$ versus $T^*$. (c) $\langle R_{22}^a\rangle$ and $\langle R_{11}^s\rangle$ versus $T^*$.

The nature of these phase transitions is of interest. To determine the order of the transitions, we have performed calculations at temperatures close to them. If an order parameter vanished discontinuously over a series of temperatures near the transition temperature, the phase transition was assigned as first order. If the order parameter vanished continuously, then it was taken to be second order [30]. Our calculations show that the transition of $NC_{2h}$ to $ND_{2h}$ was second order provided that $\lambda_{11}$ was lower than 0.14 which is approximately the tricritical point, while the phase transition from $NC_{2h}$ to $ND_{2h}$ is found to become first order in the range $0.14 \leq \lambda_{11} < 0.20$ of the scaled coupling parameter. The results also showed that the phase transition from $NC_{2h}$ to $NU$ is always first order. This behaviour is significantly different from that predicted by Model 1 where $\lambda_s = 0.2$ ($\lambda_s$ in ref. [30] is comparable to $\lambda_{22}$ in this study). Here the phase transition from $NC_{2h}$ to $ND_{2h}$ had been predicted to be always second order [30]. In addition, Model 1 also predicted that the phase transition from $NC_{2h}$ to $NU$ only occurred at one point when $\lambda_a = 0.2$ ($\lambda_a$ in ref. [30] is comparable to $\lambda_{11}$ in this study). For $\lambda_a > 0.2$, Model 1 predicts a phase transition from $NC_{2h}$ to $ND_{2h}(\perp)$, followed by a transition from $ND_{2h}(\perp)$ to either $NU$ or $I$ (see Fig. 3 (d)). The tricritical point for the model was found to be located on the $ND_{2h}(\perp)$ - $NU$ transition line where $\lambda_a$ is about 0.31; this is considerably higher than the value of the tricritical point for Model 2.

As we have seen the scaled coupling parameters $\lambda_{11}$ and $\lambda_{22}$ can be related to the geometrical properties of the molecule itself. One approach that can be used to establish this relationship is through the excluded volume methodology suggested by Straley [19]; although his approach is not strictly analytical even for orthorhombic shaped molecules. It is certainly not feasible for the monoclinic shapes. Another problem for this approach is that, the monoclinic shapes cannot be easily generalised as those with orthorhombic shape. So it is difficult to establish whether the relationship of $\lambda_{11}$ and $\lambda_{22}$ in Model 2 to the $\lambda_a$ and $\lambda_s$ in Model 1, is exact or not. Never the less, qualitatively $\lambda_{22}$ gives the degree of deviation of the molecular shape from rod-like to orthorhombic while $\lambda_{11}$ indicates how strongly the molecule deviates from orthorhombic to monoclinic symmetry.

The $ND_{2h}(\parallel)$ phase exhibited by Model 1 has a biaxial $D_{2h}$ arrangement where the minor axes are parallel to each other (see Fig. 4(b)), which is equivalent to the $ND_{2h}$ phase of Model 2. For the $ND_{2h}(\perp)$ phase (see Fig. 4(a)), the biaxial $D_{2h}$ arrangement is such that the minor axes are orthogonal to each other. Our Model 2 does not give this type of biaxial phase because one of the minor axes in Model 1 has become the principal axis in Model 1, which by the Sonnet *et al.*[33] parameterization scheme, the principal axes of molecules in the nematic phase always tend to be parallel to each other in the $ND_{2h}$ phase, hence the $ND_{2h}(\perp)$ is not able to form in Model 2. The results given in Fig. 3 suggest that the phase behaviour predicted by Model 1 are similar to those for our model as long as the value chosen for $\lambda_s$ was not in the region where the $ND_{2h}(\perp)$ phase would appear.

## VI. CONCLUSION

Generic models of mesogenic molecules with $C_{2h}$ point group symmetry which differ from Model 1 developed by Luckhurst *et al.*[30] have been formulated based on different orientations of the $C_2$ molecular axes with respect to the molecular long axis. These models have resulted in different identifications of the dominant second rank orientational order parameters and the coupling parameters, where at one point, all of the intermolecular coefficients become real. These models, based on the parameterization suggested by Sonnet *et al.* [33] and used by Luckhurst *et al.* [30], can be employed to describe the anisotropic interactions between molecules having a monoclinic shape with their long molecular axes orthogonal to the $C_2$ symmetry axis, in contrast Model 1 describes the case where the long axis is parallel to the $C_2$ symmetry axis. Based on these models, the molecular field theory for their nematic phases has been developed. This theory contains nine non-zero second rank order parameters and six non-zero coupling parameters which makes the numerical solution of the theory a formidable task. Of the nine order parameters just three are dominant and so are retained. The phase behaviour of Model 2 for $\lambda_{22} = 0.2$, 0.3 and 0.4 with a range of values for $\lambda_{11}$ have been calculated and shown as a phase map. We find that Models 2 and 3 we have proposed are equivalent to each other in term of their phase behaviour when both share the same value of $\lambda_{11}$ and $\lambda_{22}$. This similarity can be demonstrated analytically under the theoretical frame work of the molecular field theory. Never the less, even when the two models shared the same values of $\lambda_{11}$ and $\lambda_{22}$, both models can correspond to different shapes. In marked contrast to Model 1 [30], our two models do not exhibit the $ND_{2h}(\perp)$ phase exhibited by this model. This striking difference in phase behaviour occurs because for the $ND_{2h}(\perp)$ phase of Model 1 the long axes are anti-parallel and so the configuration is stable.

However, if Model 2 or 3 were to form the novel $ND_{2h}(\perp)$ phase then with the $C_2$ symmetry axis orthogonal to the long axis in the molecule the long axes in neighbouring molecules would also have to be orthogonal which would consequently destabilise the phase. Our models offer an additional and simple way to describe the monoclinic biaxial phase, which was appreciated but not considered by the previous study [30].

**ACKNOWLEDGEMENTS**

The Authors would like to thank the University of Malaya (RG072-09AFR) and the Ministry of Higher Education UM.C/625/1/HIR/MOHE/05 for various grants which supported this project. HSN also acknowledges a grant from Emeritus Professor G. R. Luckhurst which enabled him to spend six months at the University of Southampton. Generous allocations of computing resources from the University of Malaya and the University of Southampton are acknowledged.

## APPENDIX

Our aim here is to show why and under what conditions the phase behaviour of Models 2 and 3 should be equivalent. The scaled potential of mean torque for Model 2 is

$$U^2(\Omega) = -\left[\langle R_{00}\rangle R_{00}(\Omega) + 2\lambda_{22}\langle R_{22}\rangle R_{22}(\Omega) + 2\lambda_{11}\langle R_{11}^s\rangle R_{11}^s(\Omega)\right], \quad (A.1)$$

while for Model 3 it is

$$U^3(\Omega) = -\left[\langle R_{00}\rangle R_{00}(\Omega) + 2\lambda_{22}\langle R_{22}\rangle R_{22}(\Omega) + 2\lambda_{11}\langle R_{11}^a\rangle R_{11}^a(\Omega)\right]. \quad (A.2)$$

{The superscripts 1 and 2 on the U need to be replaced by 2 and 3, respectively. In addition in these two equations the terms in $\lambda_{11}$ and $\lambda_{22}$ should be exchanged as well as in Eqs (A.7), (A.8) and (A.9).

Here the angular functions associated with the dominant order parameters (see Table 1) are

$$\begin{aligned}
R_{00}(\Omega) &= \frac{1}{2}(3\cos^2\beta - 1),\\
R_{11}^s(\Omega) &= \cos\alpha\cos\beta\cos\gamma - \cos 2\beta\sin\alpha\sin\gamma,\\
R_{11}^a(\Omega) &= \cos\alpha\cos 2\beta\cos\gamma - \cos\beta\sin\alpha\sin\gamma,\\
R_{22}(\Omega) &= \left(\frac{1+\cos^2\beta}{2}\right)\cos 2\alpha\cos 2\gamma - \cos\beta\sin 2\alpha\sin 2\gamma.
\end{aligned} \quad (A.3)$$

From the trigonometric identities

$$\cos(x \pm \frac{\pi}{2}) = \mp \sin x,$$

$$\cos[2(x \pm \frac{\pi}{2})] = \cos(2x \pm \pi) = \cos 2x,$$

$$\sin(x \pm \frac{\pi}{2}) = \pm \cos x,$$  (A.4)

$$\sin[2(x \pm \frac{\pi}{2})] = \sin(2x \pm \pi) = -\sin 2x.$$

the transformations $\alpha \to \alpha + \frac{\pi}{2}$ and $\gamma \to \gamma - \frac{\pi}{2}$ ( or $\alpha \to \alpha - \frac{\pi}{2}$ and $\gamma \to \gamma + \frac{\pi}{2}$ ), leave $R_{00}(\Omega)$ and $R_{22}(\Omega)$ unchanged in Eq(A.3) while the two terms $R_{11}^s(\Omega)$ and $R_{11}^a(\Omega)$, would be exchanged i.e. $R_{11}^s(\Omega) \leftrightarrow R_{11}^a(\Omega)$.

Based on these relations, we now show that, the two models are equivalent within the theoretical frame work of the molecular field theory. We start with the molecular field theory for Model 2 and later show that this model is equivalent to that for Model 3.

The partition function for Model 2, $Q$, is

$$Q = \int \exp(-U^2(\Omega)/T^*) d\Omega.$$  (A.5)

With this partition function the order parameters predicted by Model 2 can be written as

$$\langle R_{00} \rangle = \frac{1}{Q} \int R_{00}(\Omega) \exp(-U^2(\Omega)/T^*) d\Omega,$$

$$\langle R_{22} \rangle = \frac{1}{Q} \int R_{22}(\Omega) \exp(-U^2(\Omega)/T^*) d\Omega,$$  (A.6)

$$\langle R_{11}^s \rangle = \frac{1}{Q} \int R_{11}^s(\Omega) \exp(-U^2(\Omega)/T^*) d\Omega,$$

From the partition function, we observe that

$$Q = \int \exp(-U^2(\Omega)/T^*) d\Omega$$
$$= \iiint \exp(-[\langle R_{00} \rangle R_{00}(\Omega) + 2\lambda_{22} \langle R_{22} \rangle R_{22}(\Omega) + 2\lambda_{11} \langle R_{11}^s \rangle R_{11}^s(\Omega)]/T^*) d\Omega$$
$$= \iiint \exp(-[\langle R_{00} \rangle R_{00}(\Omega') + 2\lambda_{22} \langle R_{22} \rangle R_{22}(\Omega') + 2\lambda_{11} \langle R_{11}^s \rangle R_{11}^s(\Omega')]/T^*) d\Omega'.$$

(A.7)

where $\Omega' = \left(\alpha + \frac{\pi}{2}, \beta, \gamma + \frac{\pi}{2}\right)$. Further manipulation of Eq.(A.7), based on the earlier observation made in Eqs.(A.3), we obtain

$$Q = \iiint \exp\left(-\left[\langle R_{00}\rangle R_{00}(\Omega') + 2\lambda_{22}\langle R_{22}\rangle R_{22}(\Omega') + 2\lambda_{11}\langle R_{11}^s\rangle R_{11}^s(\Omega')\right]/T^*\right) d(\alpha + \frac{\pi}{2}) d\cos\beta\, d(\gamma + \frac{\pi}{2})$$
$$= \iiint \exp\left(-\left[\langle R_{00}\rangle R_{00}(\Omega) + 2\lambda_{22}\langle R_{22}\rangle R_{22}(\Omega) + 2\lambda_{11}\langle R_{11}^s\rangle R_{11}^a(\Omega)\right]/T^*\right) d\alpha\, d\cos\beta\, d\gamma.$$

(A.8)

We denote the new expression in the exponent in Eq.(A.8) as $U'(\Omega)$, i. e.

$$U'(\Omega) = \left[\langle R_{00}\rangle R_{00}(\Omega) + 2\lambda_{22}\langle R_{22}\rangle R_{22}(\Omega) + 2\lambda_{11}\langle R_{11}^s\rangle R_{11}^a(\Omega)\right].$$ (A.9)

With the same process used in Eqs.(A.7) and Eq.(A.8) we obtain,

$$\langle R_{00}\rangle = \frac{1}{Q}\int R_{00}(\Omega) \exp\left(-U'(\Omega)/T^*\right) d\Omega,$$
$$\langle R_{22}\rangle = \frac{1}{Q}\int R_{22}(\Omega) \exp\left(-U'(\Omega)/T^*\right) d\Omega, \quad (A.10)$$
$$\langle R_{11}^s\rangle = \frac{1}{Q}\int R_{11}^a(\Omega) \exp\left(-U'(\Omega)/T^*\right) d\Omega.$$

Since,

$$\frac{1}{Q}\int R_{11}^a(\Omega) \exp\left(-U'(\Omega)/T^*\right) d\Omega = \langle R_{11}^a\rangle, \quad (A.11)$$

and in molecular field theory $\langle R_{11}^s\rangle$ can be regarded as a parameter given by the self-consistency equation in Eq.(A.6), we can redefine or replace the parameter $\langle R_{11}^s\rangle$ by $\langle R_{11}^a\rangle$. With this replacement we have shown that in terms of molecular field theory, given the same values of the scaled coefficients $\lambda_{11}$ and $\lambda_{22}$, Models 2 and 3 will have exactly the same phase behaviour.